\documentclass[twocolumn,pra,showpacs,showkeys]{revtex4}

\usepackage{dcolumn}
\usepackage{bm}

\def\etal{\emph{et~al.}}
\newcommand{\eref}[1]{(\ref{#1})}
\newcommand{\Eref}[1]{Eq.~(\ref{#1})}
\newcommand{\tref}[1]{Table~\ref{#1}}
\newcommand{\rtw}{\rightarrow}
\newcommand{\cm}{cm$^{-1}$}

\begin{document}

\title{Improved calculation of relativistic shift and isotope shift in Mg~I}

\author{J. C. Berengut}
\email{jcb@phys.unsw.edu.au}
\affiliation{School of Physics, University of New South Wales, Sydney 2052, Australia}
\author{V. V. Flambaum}
\affiliation{School of Physics, University of New South Wales, Sydney 2052, Australia}
\author{M. G. Kozlov}
\affiliation{Petersburg Nuclear Physics Institute, Gatchina, 188300, Russia}

\date{1 July 2005}

\begin{abstract}

We present an \emph{ab initio} method of calculation of isotope shift
and relativistic shift in atoms with a few valence electrons. It is based on an energy calculation involving combination of the configuration interaction method and many-body perturbation theory. This work is motivated by analyses of quasar absorption spectra that suggest that the fine structure constant $\alpha$ was smaller at an early epoch. Relativistic shifts are needed to measure this variation of $\alpha$, while isotope shifts are needed to resolve systematic effects in this study. The isotope shifts can also be used to measure isotopic abundances in gas clouds in the early universe, which are needed to study nuclear reactions in stars and supernovae and test models of chemical evolution. This paper shows that isotope shift in magnesium can be calculated to very high precision using our new method.

\end{abstract}

\pacs{06.20.-f, 31.30.Gs, 31.25.Jf}
\keywords{isotope shift; mass shift; alkaline-earth; magnesium}

\maketitle

\section{\label{sec:intro} Introduction}

The motivation for this work comes from recent studies of quasar
absorption spectra designed to probe $\alpha$ in the distant past.
Atomic transition frequencies depend on $\alpha$, and by comparing
frequencies on Earth with those in quasar absorption spectra, one
can deduce whether or not $\alpha$ was different in the early universe.
While some studies have revealed a significant deviation from zero
(Refs.~\cite{webb99prl,webb01prl,murphy01mnrasA,webb03ass}) other
groups using a different telescope have not
(Refs.~\cite{quast04aap,srianand04prl}).

One of the possible major sources of
systematic effects in these studies is that the isotopic abundance
ratios in gas clouds in the early universe could be 
very different to those on Earth. A ``conspiracy'' of several isotopic
abundances may provide an alternative explanation for the observed 
variation in spectra \cite{murphy01mnrasB,murphy03ass,murphy03mnras}.
In order to test this possibility
it is necessary to have accurate values for the isotope shift (IS) for
the relevant atomic transitions \cite{kozlov04pra}.
Experimental data are available for
very few of them; therefore, accurate calculations are needed
in order to make the most comprehensive analysis possible.

The need for accurate isotope shifts is further motivated by a wish
to study isotopic evolution in the universe. The isotopic abundances
of gas clouds may be measured independently of a variation in $\alpha$
\cite{kozlov04pra}. This is important for testing models of nuclear reactions in
stars and supernovae and of the chemical evolution of the universe.

Previously we have calculated isotope shift in atoms with one valence
electron using many-body perturbation theory (MBPT) \cite{berengut03pra},
and for neutral magnesium using the configuration interaction (CI) method \cite{berengut04praA}.
Both of these papers used the finite-field scaling
method for the perturbation. In a sense our method combines
these two methods by including core-valence correlations into our CI method
using MBPT. Magnesium is one of the simplest and
well studied two-valence-electron atoms. Because of that it is often used as a
test ground for different methods of atomic calculations. In this
paper we show that we can calculate the isotope shift of some
magnesium transitions for which experimental values are available.

\section{\label{sec:method} Method}

The isotope shifts of atomic transition frequencies come from two
sources: the finite size of the nuclear charge distribution (the
``volume'' or ``field'' shift), and the finite mass of the nucleus
(see, e.g. \cite{sobelman72book}). The energy shift due to recoil of the
nucleus is $(1/2M)\bm{p}_N^2~=~(1/2M)(\Sigma \bm{p}_i)^2$. 
Furthermore this ``mass shift'' is traditionally divided
into the normal mass shift (NMS) and the specific mass shift (SMS). The
normal mass shift is given by the operator $(1/2M)\Sigma \bm{p}_i^2$, 
which is easily calculated from the transition frequency. The
SMS operator is $(1/M)\Sigma_{i<j}(\bm{p}_i \cdot \bm{p}_j)$ which is
difficult to evaluate accurately.

The shift in energy of any transition in an isotope with mass number
$A'$ with respect to an isotope with mass number $A$ can be expressed
as
\begin{equation}
\label{eq:is}
\delta \nu^{A', A} = \left( k_{\rm NMS} + k_{\rm SMS} \right)  \left(
    \frac{1}{A'} - \frac{1}{A} \right) + F \delta \langle r^2 \rangle
    ^{A', A} \ ,
\end{equation}
where the normal mass shift constant is
\begin{equation}
k_{\rm NMS} = -\frac{\nu}{1823}
\end{equation}
and $\langle r^2 \rangle$ is the mean square nuclear radius. The value
1823 refers to the ratio of the atomic mass unit to the electron mass.

In this paper we develop a method for calculating the specific mass
shift $k_{\rm SMS}$ for atoms with several valence electrons.
It is worth noting that in this paper we use the
convention $\delta \nu^{A', A}~=~\nu^{A'} - \nu^{A}$.

Following our previous work, we are looking for an ``all orders'' method of
calculation. Again we have found that the finite-field scaling method
is very useful in this respect. The rescaled SMS operator is added to
the many-particle Hamiltonian
\begin{eqnarray}
\label{eq:H_sms}
H_{\lambda} = H_0 + \lambda H_{\rm SMS} = H_0 
+ \lambda \sum_{i<j}\bm{p}_i \cdot \bm{p}_j.
\end{eqnarray}
The eigenvalue problem for the new Hamiltonian is solved for various
$\lambda$, and then we recover the specific mass shift constant as
\begin{equation}
k_{\rm SMS} = \lim_{\lambda \rtw 0} \frac{dE}{d\lambda}.
\end{equation}
The operator \eref{eq:H_sms} has the same symmetry and structure as
the initial Hamiltonian $H_0$ (see the Appendix, Ref.~\cite{berengut03pra}).

We have also calculated the relativistic shift for transitions in magnesium using
a similar method. These are needed in order to measure variation of $\alpha$
\cite{webb99prl,webb01prl,murphy01mnrasA,webb03ass,quast04aap,srianand04prl}.
The dependence of transition frequencies on $\alpha$ can be
expressed as
\begin{equation}
\label{eq:alpha}
\omega = \omega_0 + q x,
\end{equation}
where $x = \left( \alpha / \alpha_0 \right)^2 - 1$, $\alpha_0$ is the laboratory value
of the fine-structure constant, and $\omega$ and $\omega_0$ are the frequencies of the
transition in the absorption clouds and in the laboratory, respectively.
We vary $\alpha$ directly in codes and hence calculate $q$. This method is described
in detail in our earlier works (Refs.~\cite{dzuba99prl,dzuba02pra,berengut04praB}).

To calculate the energy, we use a combination of configuration interaction (CI) and
many-body perturbation theory (MBPT), as was done in Refs.~\cite{dzuba96pra,dzuba98pra}.
We generate a basis set that includes the core and valence orbitals and a number of
virtual orbitals. Then we do the full configuration interaction calculation in the
frozen-core approximation. Core-valence correlation effects are taken into account using MBPT by altering the Coulomb integrals
($\tilde Q = 1/\left| \bm{r}_i - \bm{r}_j \right|
             + \lambda \bm{p}_i \cdot \bm{p}_j$)
in the CI calculation. The MBPT operator
is separated into two parts: $\Sigma_1$ and $\Sigma_2$, which include all
one-valence-electron diagrams and two-valence-electron diagrams, respectively.
The operators are calculated to
second-order, which was shown to be sufficient for single-valence-electron atoms when
used with a finite-field method \cite{berengut03pra}.

\section{Calculation and results}

The CI part of the calculation is very similar to the large basis set calculation in Ref.~\cite{berengut04praA}. We first solved the Dirac-Fock equations for the core electrons; we use the $V^{N-2}$ approximation, so the core includes the orbitals $1s_{1/2}$, $2s_{1/2}$, $2p_{1/2}$, and $2p_{3/2}$.
For valence and virtual orbitals we used a basis formed by diagonalizing the Dirac-Fock operator on the basis set of B-splines and excluding orbitals with high energy (for a description of this method as applied in atomic physics, see, e.g.~Refs.~\cite{johnson86prl,johnson87pra,johnson88pra}).

The full two-electron CI uses the basis $17spdf$, which includes the orbitals
$1 - 17s_{1/2}$, $2 - 17p_j$, $3 - 17d_j$, and $4 - 17f_j$.
It is very close to the saturation of the valence CI. The MBPT basis can be larger, since the calculation merely involves a summation over the virtual orbitals; we have therefore used the basis $32spdfg$ for this part of the calculation, which is essentially complete.

\tref{tab:energy} shows that our \emph{ab initio} calculation of the
spectrum is within 0.4\% of the experimental spectrum for all considered levels.
The relativistic shifts ($q$-values) are presented in \tref{tab:relshift}.
\tref{tab:SMSshifts} presents the resulting SMS level shift constants, $k_{\rm SMS}$ of \Eref{eq:is}.
In each table we present results of the pure CI calculation
(which agree with our previous calculation Ref.~\cite{berengut04praA}),
as well as calculations including $\Sigma_1$ only,
and both $\Sigma_1$ and $\Sigma_2$.

\begin{table}
\caption{Energy of Mg~I levels relative to the ground state ($3s^2\ ^1S_0$).}
\label{tab:energy}
\begin{tabular}{lcccc}
\hline \hline
  & \multicolumn{4}{c}{Energy (\cm)} \\
\multicolumn{1}{c}{Level} & \multicolumn{1}{c}{Experiment} & \multicolumn{1}{c}{CI}
  & \multicolumn{1}{c}{CI + $\Sigma_{1}$}
  & \multicolumn{1}{c}{CI + $\Sigma_{1}$ + $\Sigma_{2}$} \\
\hline
$3s3p\ ^3\!P_0^o$ &$ 21850 $&$ 20910 $&$ 21676 $&$ 21772 $\\
$3s3p\ ^3\!P_1^o$ &$ 21870 $&$ 20930 $&$ 21698 $&$ 21794 $\\
$3s3p\ ^3\!P_2^o$ &$ 21911 $&$ 20971 $&$ 21742 $&$ 21837 $\\
$3s3p\ ^1\!P_1^o$ &$ 35051 $&$ 34491 $&$ 35474 $&$ 35050 $\\
$3s4s\ ^3\!S_1$   &$ 41197 $&$ 40406 $&$ 41469 $&$ 41126 $\\
$3s4s\ ^1\!S_0$   &$ 43503 $&$ 42667 $&$ 43744 $&$ 43431 $\\
$3s3d\ ^1\!D_2$   &$ 46403 $&$ 45123 $&$ 46475 $&$ 46306 $\\
$3s4p\ ^3\!P_0^o$ &$ 47841 $&$ 46919 $&$ 48079 $&$ 47756 $\\
$3s4p\ ^3\!P_1^o$ &$ 47844 $&$ 46923 $&$ 48082 $&$ 47760 $\\
$3s4p\ ^3\!P_2^o$ &$ 47851 $&$ 46929 $&$ 48090 $&$ 47767 $\\
$3s3d\ ^3\!D_3$   &$ 47957 $&$ 46973 $&$ 48227 $&$ 47880 $\\
$3s3d\ ^3\!D_2$   &$ 47957 $&$ 46973 $&$ 48227 $&$ 47879 $\\
$3s3d\ ^3\!D_1$   &$ 47957 $&$ 46973 $&$ 48227 $&$ 47879 $\\
$3s4p\ ^1\!P_1^o$ &$ 49347 $&$ 48490 $&$ 49672 $&$ 49277 $\\
$3s4d\ ^1\!D_2$   &$ 53135 $&$ 52041 $&$ 53337 $&$ 53037 $\\
$3s4d\ ^3\!D_1$   &$ 54192 $&$ 53243 $&$ 54486 $&$ 54111 $\\
$3p^2\ ^3\!P_0$   &$ 57813 $&$ 56182 $&$ 58003 $&$ 57706 $\\

\hline\hline
\end{tabular}
\end{table}

\begin{table}
\caption{Relativistic shift of Mg~I transitions relative to the ground state ($3s^2\ ^1S_0$).}
\label{tab:relshift}
\begin{tabular}{lcccc}
\hline \hline
Upper & Energy & \multicolumn{3}{c}{$q$ (\cm)} \\
Level & (\cm)  & CI & CI + $\Sigma_{1}$ & CI + $\Sigma_{1}$ + $\Sigma_{2}$ \\
\hline
$3s3p\ ^3\!P_0^o$ &$ 21850 $&$  61 $&$  66 $&$  68 $\\
$3s3p\ ^3\!P_1^o$ &$ 21870 $&$  81 $&$  88 $&$  89 $\\
$3s3p\ ^3\!P_2^o$ &$ 21911 $&$ 122 $&$ 132 $&$ 133 $\\
$3s3p\ ^1\!P_1^o$ &$ 35051 $&$  86 $&$  94 $&$  94 $\\
$3s4s\ ^3\!S_1$   &$ 41197 $&$  55 $&$  61 $&$  61 $\\
$3s4s\ ^1\!S_0$   &$ 43503 $&$  60 $&$  65 $&$  66 $\\
$3s3d\ ^1\!D_2$   &$ 46403 $&$ 117 $&$ 123 $&$ 122 $\\
$3s4p\ ^3\!P_0^o$ &$ 47841 $&$  67 $&$  73 $&$  73 $\\
$3s4p\ ^3\!P_1^o$ &$ 47844 $&$  70 $&$  76 $&$  77 $\\
$3s4p\ ^3\!P_2^o$ &$ 47851 $&$  77 $&$  83 $&$  84 $\\
$3s3d\ ^3\!D_3$   &$ 47957 $&$  79 $&$  85 $&$  86 $\\
$3s3d\ ^3\!D_2$   &$ 47957 $&$  79 $&$  85 $&$  86 $\\
$3s3d\ ^3\!D_1$   &$ 47957 $&$  79 $&$  85 $&$  86 $\\
$3s4p\ ^1\!P_1^o$ &$ 49347 $&$  80 $&$  86 $&$  87 $\\
$3s4d\ ^1\!D_2$   &$ 53135 $&$  94 $&$ 101 $&$ 102 $\\
$3s4d\ ^3\!D_1$   &$ 54192 $&$  73 $&$  79 $&$  80 $\\
$3p^2\ ^3\!P_0$   &$ 57813 $&$ 198 $&$ 214 $&$ 214 $\\

\hline\hline
\end{tabular}
\end{table}

\begin{table}
\caption{Calculations of the specific mass shift constants $k_{\rm
SMS}$ for Mg~I transitions to the ground state (in $ \rm GHz \cdot amu$).}
\label{tab:SMSshifts}
\begin{tabular}{lcccc}
\hline \hline
Upper & \multicolumn{1}{c}{Energy} & \multicolumn{3}{c}{$k_{\rm SMS}$ (GHz.amu)} \\
Level & \multicolumn{1}{c}{(\cm)}
  & \multicolumn{1}{c}{CI}
  & \multicolumn{1}{c}{CI + $\Sigma_{1}$}
  & \multicolumn{1}{c}{CI + $\Sigma_{1}$ + $\Sigma_{2}$} \\
\hline
$3s3p\ ^3\!P_0^o$ &$ 21850 $&$ -378 $&$ -487 $&$ -492 $\\
$3s3p\ ^3\!P_1^o$ &$ 21870 $&$ -377 $&$ -486 $&$ -491 $\\
$3s3p\ ^3\!P_2^o$ &$ 21911 $&$ -375 $&$ -485 $&$ -489 $\\
$3s3p\ ^1\!P_1^o$ &$ 35051 $&$  231 $&$  120 $&$  134 $\\
$3s4s\ ^3\!S_1$   &$ 41197 $&$   43 $&$  -59 $&$  -49 $\\
$3s4s\ ^1\!S_0$   &$ 43503 $&$   13 $&$  -94 $&$  -85 $\\
$3s3d\ ^1\!D_2$   &$ 46403 $&$ -345 $&$ -500 $&$ -477 $\\
$3s4p\ ^3\!P_0^o$ &$ 47841 $&$  -17 $&$ -136 $&$ -126 $\\
$3s4p\ ^3\!P_1^o$ &$ 47844 $&$  -16 $&$ -136 $&$ -126 $\\
$3s4p\ ^3\!P_2^o$ &$ 47851 $&$  -16 $&$ -136 $&$ -126 $\\
$3s3d\ ^3\!D_3$   &$ 47957 $&$   52 $&$  -87 $&$  -77 $\\
$3s3d\ ^3\!D_2$   &$ 47957 $&$   52 $&$  -87 $&$  -77 $\\
$3s3d\ ^3\!D_1$   &$ 47957 $&$   52 $&$  -87 $&$  -77 $\\
$3s4p\ ^1\!P_1^o$ &$ 49347 $&$    5 $&$ -120 $&$ -108 $\\
$3s4d\ ^1\!D_2$   &$ 53135 $&$ -100 $&$ -246 $&$ -239 $\\
$3s4d\ ^3\!D_1$   &$ 54192 $&$   32 $&$  -99 $&$  -88 $\\
$3p^2\ ^3\!P_0$   &$ 57813 $&$ -225 $&$ -469 $&$ -464 $\\

\hline\hline
\end{tabular}
\end{table}

It is worth noting a few points. Firstly, the core-valence effects, included
using MBPT, make little difference to the $q$-values (less than 10\%).
This again justifies the fact that in previous works for atoms with several
valence electrons these have either been
neglected entirely or included using a simple fitting procedure based on the
polarisability of the core \cite{berengut04praB}.

The core-valence effects are much more important for the SMS calculation. In
particular the single-valence-electron diagrams (included in $\Sigma_1$) can improve
accuracy drastically in cases where the pure CI method is not very good.
Although $\Sigma_2$ is important for energy calculation, it appears to make little
difference to $k_{\rm SMS}$. This is easily understood since the most important
two-body diagram (the direct diagram, corresponding to the screening of the electron-electron
interaction by the core electrons) makes no contribution to the SMS. The exchange diagrams in
$\Sigma_2$ do have an effect, but this is much smaller than the one-body contribution.

In \tref{tab:comparisons} we compare experimental and calculated frequency
shifts between isotopes $^{26}{\rm Mg}$ and $^{24}{\rm Mg}$ ($\delta \nu^{26, 24}$).
We compare the SMS part only, which is extracted from experiment by subtracting the NMS.
We have ignored the volume shift for simplicity; it is approximately
\hbox{20-30~MHz}, which
is less than the experimental uncertainty in most cases and is of the order of the
error in our SMS calculations. Furthermore despite numerous convergent
theoretical calculations of the field shift parameter ($F$ of \Eref{eq:is}),
our knowledge of the field shift in magnesium
is limited by our lack of knowledge of the change in mean-square nuclear radius
$\delta \left< r^2 \right>^{26, 24}$.

Also presented in \tref{tab:comparisons}, for a theoretical comparison, are
the results of Veseth (Ref.~\cite{veseth87jpb}, 1987) and
J\"onsson \etal~(Ref.~\cite{jonsson99jpb}, 1999).
Veseth used non-relativistic many-body perturbation theory within the algebraic
approximation to calculate the isotope shift to third order for some transitions.
J\"onsson \etal~used a non-relativistic multiconfiguration Hartree-Fock approach,
which allowed for both core and valence excitations in the CI.

\begin{table*}[htb]
\caption{Comparison of our calculated SMS with that extracted from experiment
for several transitions (in MHz). The isotope shifts are between 
$^{26}{\rm Mg}$ and $^{24}{\rm Mg}$.
Also presented are the results of Refs.~\cite{veseth87jpb}
and~\cite{jonsson99jpb} for theoretical comparison.
We have neglected the field shift; it is of the order of 20-30 MHz.}
\label{tab:comparisons}
\begin{tabular}{lccccccccc}
\hline \hline
  &\multicolumn{1}{c}{$\lambda$}
  &\multicolumn{1}{c}{IS(expt.)} 
  &\multicolumn{1}{c}{NMS}
  &\multicolumn{6}{c}{SMS} \\
\multicolumn{1}{l}{Transition}
  &\multicolumn{1}{c}{(\AA)}
  & & &\multicolumn{1}{c}{CI}
  &\multicolumn{1}{c}{CI + $\Sigma_1$}
  &\multicolumn{1}{c}{CI + $\Sigma_1$ + $\Sigma_2$}
  &\multicolumn{1}{c}{Expt.}
  &\multicolumn{1}{c}{Ref.~\cite{veseth87jpb}}
  &\multicolumn{1}{c}{Ref.~\cite{jonsson99jpb}} \\
\hline
$3s^2\ ^1\!S_0 \rtw 3s3p\ ^3\!P_1^o$ & 4572 & 2683(0)\footnotemark[1]
    & 1153 &  1208  &  1559 &  1573 & 1530 & 1378 & 1666 \\
$3s^2\ ^1\!S_0 \rtw 3s3p\ ^1\!P_1^o$ & 2853 & 1412(21)\footnotemark[2]
    & 1848 &  -740  &  -383 &  -428 & -436 &      & -409 \\
                                     &      & 1390(31)\footnotemark[3]
    &      &        &       &       & -458 & & \\
$3s3p\ ^3\!P_0^o \rtw 3s4s\ ^3\!S_1$ & 5169 & -396(6)\footnotemark[4]
    & 1020 & -1347 & -1371 & -1419 & -1416 & & \\
$3s3p\ ^3\!P_1^o \rtw 3s4s\ ^3\!S_1$ & 5174 & -390(5)\footnotemark[4]
    & 1019 & -1345 & -1369 & -1416 & -1409 & & \\
$3s3p\ ^3\!P_2^o \rtw 3s4s\ ^3\!S_1$ & 5185 & -390(7)\footnotemark[4]
    & 1017 & -1339 & -1363 & -1411 & -1407 & & \\
$3s3p\ ^3\!P_1^o \rtw 3p^2\ ^3\!P_0$ & 2782 & 1810(80)\footnotemark[5]
    & 1895 & -486  &  -56  &   -86 &   -85 & & \\
$3s3p\ ^3\!P_0^o \rtw 3s3d\ ^3\!D_1$ & 3830 & 60(15)\footnotemark[2]
    & 1376 & -1377 & -1283 & -1329 & -1316 & -1269 & \\
$3s3p\ ^3\!P_1^o \rtw 3s3d\ ^3\!D_{1,2}$ & 3833 & 61(3)\footnotemark[2]
    & 1375 & -1374 & -1280 & -1326 & -1314 & & \\
$3s3p\ ^3\!P_2^o \rtw 3s3d\ ^3\!D_{1,2,3}$ & 3839 & 58(4)\footnotemark[2]
    & 1373 & -1368 & -1274 & -1321 & -1315 & & \\
$3s3p\ ^3\!P_1^o \rtw 3s4d\ ^3\!D_1$ & 3094 & 420(20)\footnotemark[5]
    & 1704 & -1309 & -1241 & -1291 & -1284 & & \\
$3s3p\ ^1\!P_1^o \rtw 3s4d\ ^1\!D_2$ & 5530 & 2107(15)\footnotemark[4]
    & 953  & 1059  &  1173 &  1195 &  1154 & & \\

\hline\hline
\end{tabular}
\footnotetext[1]{Sterr \etal, 1993 \cite{sterr93apb}}
\footnotetext[2]{Hallstadius, 1979 \cite{hallstadius79zpa}}
\footnotetext[3]{Le Boiteux \etal, 1988 \cite{boiteux88jpf}}
\footnotetext[4]{Hallstadius and Hansen, 1978 \cite{hallstadius78zpa}}
\footnotetext[5]{Novero \etal, 1992 \cite{novero92nc}}
\end{table*}

An under-studied transition that is seen in quasar absorption spectra is the
$2026$~\AA~line of MgI ($3s^2\ ^1\!S_0 \rtw 3s4p\ ^1\!P_1^o$).
From \tref{tab:SMSshifts}, we calculate the isotope shift
of this line as $\delta \nu^{26, 24} = 2950(50)$~MHz (the error here is
based on the absence of field shift as well as the incompleteness of saturation 
of the basis set used to calculate $k_{\rm SMS}$).

\section{Conclusion}

We have presented a method for the calculation of the isotope-shift in
many-electron atoms. It is based on the finite-field method, with an
energy calculation that combines CI for the valence electrons and
MBPT for the core-valence correlations. We have tested the method in
magnesium, and the agreement was found to be very good for all
transitions. In particular, for the purposes of resolving systematic
errors in the search for $\alpha$-variation, and for studies of
isotopic evolution of the universe, such accuracy is high enough.

We have also used the method to generate more precise values for the
relativistic shift ($q$-values). These were found to be within 10\% of those
found using previous methods, as expected.

\section{Acknowledgments}

This work is supported by the Australian Research Council,
Gordon Godfrey fund, and Russian Foundation for Basic Research,
grant No.~05-02-16914.
The authors would like to thank V. A. Dzuba for providing an updated
version of the Dzuba - Flambaum - Sushkov atomic code. We are
grateful to the APAC National Facility for providing computer time.

\bibliography{references}

\end{document}